\documentclass{article}
\usepackage{iclr2027_conference,times}
\iclrfinalcopy 
\PassOptionsToPackage{hyphens}{url}
\usepackage{hyperref}
\usepackage{url}
\usepackage{graphicx}
\usepackage{caption}
\usepackage{soul}
\usepackage{amsmath,amssymb}
\usepackage{xcolor}
\usepackage{placeins}

\title{Active Data Acquisition with Side Information via Discrete Diffusion Priors}

\author{An Vuong, Thinh Nguyen\\
Department of Electrical Engineering and Computer Science \\
Oregon State University \\
Corvallis, OR 97331, USA. \\
\texttt{\{vuonga2,thinhq\}@oregonstate.edu}}

\begin{document}
\maketitle

\begin{abstract}
Acquiring data is costly: higher measurement fidelity costs power and storage and risks collecting irrelevant content, while aggressive cost reduction can discard information that later analysis needs. We address this trade-off with an information-theoretic framework that acquires data relevant to a broad set of tasks rather than to one model. A mask policy, conditioned on side information, chooses which pixels to measure so as to maximize the mutual information between a discrete image and its partial observation under a budget; since the image entropy does not depend on the mask, this is equivalent to minimizing the conditional entropy. A frozen discrete denoising diffusion model (D3PM) supplies the posterior, and we use it in two ways: as an entropy surrogate for training a one-shot mask generator, and as the criterion for sequential greedy acquisition. The one-shot generator outperforms random masks only with care, including an unbiased gradient estimator for binary masks. With sequential acquisition, on MNIST the prior makes $8\times$ fewer errors than random at a $10\%$ budget, and on CIFAR-10 it gains $0.9$--$3.4$~dB. On fastMRI, our proposed technique using a static mask outperforms the well-known methods such as variable density and LOUPE.
\end{abstract}


\section{Introduction}

Measuring everything is expensive, and measuring too little discards information later tasks need. In budget-constrained pixel sensing, such as medical imaging, remote sensing and bandwidth-limited transmission, only a fraction of locations can be measured, and the question is which ones carry the most information about the whole picture. In real-time systems such as autonomous driving, sensor bandwidth, power and latency limit how much can be measured and processed for each frame \citep{gehrig2024lowlatency}, so choosing what to measure also reduces the computation downstream; adaptive LiDAR, for instance, samples only part of the scene \citep{bergman2020deep,pittaluga2020mems}.

\begin{figure}[t]
\centering
\includegraphics[width=\textwidth]{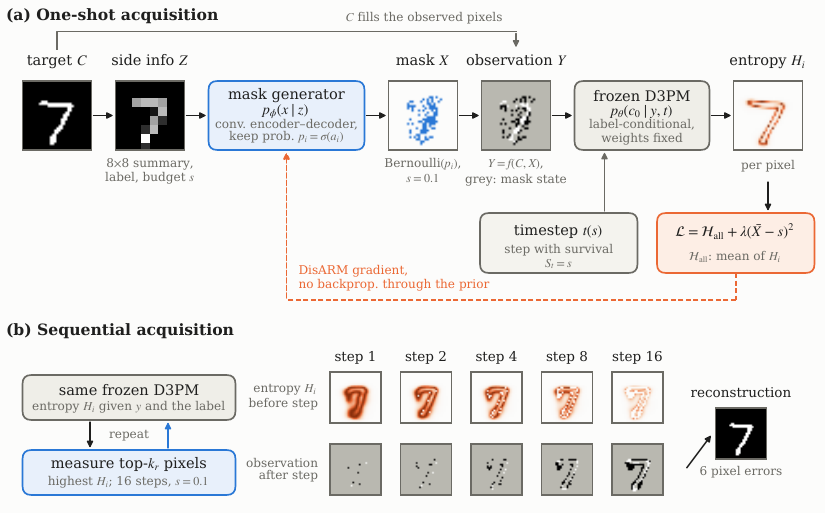}
\caption{The two uses of the prior, on one MNIST test digit at $s = 0.1$. (a) Training the one-shot mask generator. From the side information $Z$, the generator gives each pixel a keep probability $p_i$; a sampled mask $X$ yields the observation $Y$, in which unobserved pixels take the mask state (grey). The frozen D3PM, queried at the timestep $t(s)$ whose survival matches the budget, gives per-pixel entropies $H_i$, and the loss is their mean plus the budget penalty. The generator is trained with DisARM from the loss of antithetic mask pairs; no gradient passes through the prior. (b) Sequential acquisition with the same prior: starting from no observations, with only the label, each of 16 steps measures the unobserved pixels of highest entropy and updates the posterior. Insets are outputs of the trained generator and prior.}
\label{fig:pipeline}
\end{figure}

We take an information-theoretic view. Rather than tailoring acquisition to one downstream model, we score measurements by their mutual information with the target, a model-agnostic quantity \citep{cover2006elements}. A sensing action $X$ is chosen from side information $Z$, i.e., a class label, a coarse summary of the scene and a budget, to produce a measurement $Y$ with maximal information about the target $C$ (Figure~\ref{fig:pipeline}). Here the sensing action is a binary mask over pixels, the setting of our experiments. Learned masks have been trained jointly with a reconstructor \citep{bahadir2020loupe,huijben2020dps,aggarwal2020jmodl}, as sequential policies that choose the next measurement from the current reconstruction \citep{zhang2019reducing,pineda2020active,bakker2022adaptive}, and, in Bayesian experimental design, against a bound or neural estimate of the mutual information between measurements and the quantity of interest \citep{kleinegesse2020mine,foster2021dad}; \citet{vuong2025cnc} learn masks for continuous images this way, pairing a U-Net mask generator \citep{ronneberger2015unet} with MINE \citep{belghazi2018mine}. Here the target is discrete, and a frozen absorbing-state D3PM \citep{austin2021structured} supplies a surrogate for $H(C \mid Y)$.

Training a mask generator lowers the value of this surrogate objective, but a lower objective value does not by itself mean better masks. We therefore judge every mask by a criterion it was not trained on: how well the image is reconstructed from the pixels the mask observed; on accelerated MRI, this is the reconstruction error of a network that played no part in choosing the mask.

Those checks separate two ways of optimizing the same objective. In \textbf{one-shot} acquisition, a mask generator is trained against the surrogate and chooses the whole mask before anything is measured. In \textbf{sequential} acquisition, the posterior is evaluated after each round of measurements and the next pixels are chosen where it is most uncertain, similar to \cite{nolan2025ads}. For noiseless pixel observations that uncertainty is exactly the information gained, $I(C; Y_i \mid y) = H(C_i \mid y)$, so the sequential method is greedy maximization of $I(Y;C)$ itself.

\subsection{Contributions}
\begin{itemize}
\item We cast budgeted acquisition with side information as maximizing $I(Y;C)$ under a budget constraint for discrete images. The objective is task-agnostic: it values a measurement by what it reveals about the whole target, not by one reconstructor's or classifier's loss, so one mask serves many downstream tasks. The policy conditions on side information $Z$ (class label, coarse summary, budget), and a frozen D3PM, queried at a timestep calibrated to the budget, gives a tractable surrogate for the conditional entropy $H(C \mid Y)$. We measure what each part of $Z$ contributes (Table~\ref{tab:side-info}).
\item We compare two ways of using the prior. One-shot optimization of the entropy surrogate needs care: averaging over observed pixels is degenerate, since it rewards observing predictable background, and the straight-through Gumbel-Softmax gradient points away from the whole-image objective. Trained with an unbiased estimator (DisARM), the one-shot mask makes fewer errors than random masks on MNIST at every budget, but on CIFAR-10 its gain is under $1$~dB and appears only with the target's own summary. Sequential greedy information gain works better: in 16 steps it makes $8\times$ fewer errors than random on MNIST at $s = 0.1$, and on CIFAR-10 it improves on random by $0.9$--$3.4$~dB PSNR.
\item We also test on the fastMRI dataset: the one-shot surrogate disagrees with reconstruction quality, instance conditioning does not contribute meaningful improvement, mask ranking depends on the reconstructor, and a static mask optimized against a reconstructor has lower reconstruction error than variable density and LOUPE \citep{bahadir2020loupe}.
\end{itemize}

\section{Related Work}

\paragraph{Active learning and sensing.}
Active learning selects samples to label for a particular model, for instance by their mutual information with the unknown labels \citep{guo2007active}; \citet{shayovitz2023universal} minimize conditional information in a universal framework. Radar detection and classification likewise optimize for a fixed task \citep{kay1998detection,han2021radar}. Scoring a batch by the sum of its items' marginal entropies is a known failure mode, because it ignores redundancy between the selected items \citep{kirsch2019batchbald}, and our per-pixel surrogate inherits that weakness. We instead score spatial acquisition masks by mutual information under explicit budgets.

\paragraph{Prior acquisition work.}
Acquisition has been cast as Bayesian experimental design, choosing measurements to maximize their expected information about the quantity of interest \citep{kleinegesse2020mine,foster2021dad}, and, for imaging, as a sequential policy trained on reconstruction error \citep{zhang2019reducing,pineda2020active}. \citet{vuong2025cnc} pose acquisition with side information as a family of problems (P1--P4) and estimate $I(Y;C)$ with MINE on CelebA \citep{liu2015celeba}; we follow the budgeted form P4, a policy $p(x \mid z)$ under a sparsity constraint, and move to discrete states with a D3PM prior. Closest to our setting, \citet{nolan2025ads} drive active subsampling with a pre-trained diffusion model under a maximum-entropy criterion, selecting the locations whose measurements are least predictable, the direction our observed-pixel form inverts, and the direction our sequential method follows. Our method differs in using a discrete absorbing prior, in reading each pixel's information gain in closed form from one forward pass per round rather than from posterior samples, and in showing that a separate mask state matters for absorbing priors (Appendix~\ref{app:token}).

\paragraph{Discrete and continuous diffusion.}
D3PMs \citep{austin2021structured} extend denoising diffusion \citep{ho2020ddpm} to categorical data; the absorbing-state kernel matches masked prediction. We use an absorbing D3PM with a cosine schedule \citep{nichol2021improved}, built on the implementation of \citet{ryu2024d3pm}, and a DiT backbone \citep{peebles2023dit} on CIFAR-10. Masked diffusion of this kind now scales to large language models, trained on a bound on the likelihood \citep{nie2025lld}; we use the prior differently, as a frozen scorer, and our surrogate is a mean of per-pixel entropies of its $c_0$-predictor. The usual relaxation for training over binary masks, Gumbel-Softmax \citep{jang2017gumbel,maddison2017concrete}, gives a gradient that points the wrong way here, so we train with DisARM \citep{dong2020disarm}.

\paragraph{Learned sampling for MRI.}
Variable-density random sampling with a fully sampled low-frequency block is the standard Cartesian heuristic \citep{lustig2007sparse}. Learned patterns follow two lines of work: combinatorial selection, building masks greedily against reconstruction error \citep{gozcu2018learning}, and relaxation, training a probabilistic mask jointly with a reconstruction network, as in LOUPE \citep{bahadir2020loupe}. Our MRI experiments re-derive both to test whether an information-theoretic, instance-adaptive policy improves on them; it does not, and we run LOUPE itself as a baseline. \citet{bakker2022adaptive} already found adaptive acquisition policies roughly matching non-adaptive ones on multi-coil fastMRI, so our conditioning result replicates theirs in the single-coil Cartesian setting rather than establishing something new. \citet{alkan2025autosamp} maximize a variational mutual-information bound for 3D k-space design with the reconstructor trained jointly, and improve on variable density, the regime our own positive result points to.

\section{Problem Formulation}
\label{sec:formulation}

We write $C$ for the target discrete image, $X$ for the sensing action (a binary mask) chosen by a policy $p(x \mid z)$, $Y$ for the measurement, a partial observation of $C$, and $Z$ for the side information. Images are discretized to $N$ bins per channel, so $C \in \{0,\ldots,N-1\}^{C_{\mathrm{ch}} \times H \times W}$ over $C_{\mathrm{ch}}$ channels. A mask $X \in \{0,1\}^{1 \times H \times W}$ selects observed pixels, and $Y = f(C, X)$ replaces every unobserved pixel with the D3PM's absorbing token. The token has a state of its own: $0$ means unobserved and nothing else, and the data bins are $1,\ldots,N$, giving $3$ states on MNIST ($0 = $ mask, $1 = $ black, $2 = $ white) and $9$ on CIFAR-10. Were the token to share a value with the data, an observed black pixel and an unobserved pixel would be the same input to the prior; Appendix~\ref{app:token} measures what the separate state adds. Side information $Z$ is an $8 \times 8$ grayscale summary of $C$, the class label (MNIST), and a budget $s$ drawn per sample from $\mathcal{U}(s_{\min}, s_{\max})$.

\paragraph{Problem.}
\begin{equation}
\max_{p(x \mid z)} \; I(Y; C) \quad \text{s.t.} \quad \mathbb{E}\Bigl[\tfrac{1}{HW}\textstyle\sum_i X_i\Bigr] = s.
\label{eq:p4}
\end{equation}
Since $I(Y;C) = H(C) - H(C \mid Y)$ and $H(C)$ does not depend on $X$, this is equivalent to minimizing $H(C \mid Y)$. Optimal solutions of this lie at extreme points of convex feasible sets, and enumerating them is intractable in high dimensions \citep{vuong2025cnc}; we instead parameterize $p_\phi(x \mid z)$ and minimize a differentiable surrogate.

\section{Method}
\label{sec:method}

\subsection{A D3PM surrogate for \texorpdfstring{$H(C \mid Y)$}{H(C|Y)}}
A frozen absorbing-state D3PM \citep{austin2021structured} with denoiser $p_\theta(c_0 \mid c_t, t, c_{\mathrm{lbl}})$ is applied to the observation $Y$ at a timestep $t(s)$ matched to the budget. Its per-pixel entropy is
\begin{equation}
H_i = -\sum_{k=0}^{N-1} p_i(k) \log p_i(k), \qquad p_i = \mathrm{softmax}(\hat{z}_i).
\end{equation}
Subadditivity makes $\sum_i H_i$ an upper bound on the joint conditional entropy under the model. Instead of the sum, we average $H_i$ in one of two ways:

\paragraph{Observed-pixel aggregation.}
Averaging $H_i$ over the observed pixels only, our original design, measures how predictable the observed pixels are rather than $H(C \mid Y)$, so it rewards observing whatever is easiest to predict: on MNIST the background, and with a weak budget penalty a nearly empty mask. We therefore average over every pixel.

\paragraph{Whole-image aggregation.}
Averaging over every pixel,
\begin{equation}
\mathcal{H}_{\mathrm{all}} = \frac{1}{HW C_{\mathrm{ch}}}\sum_i H_i,
\end{equation}
so a mask can lower the objective's value only through what its observations add to $Y$. On MNIST, observing a background pixel leaves $Y$ unchanged and earns nothing.

The D3PM is queried at a timestep $t(s)$ matched to the budget. Each step of its forward process absorbs more pixels into the mask token, so a mask that keeps a fraction $s$ of the pixels looks like a forward-corrupted image at the step where a fraction $s$ of pixels still survive. We estimate this survival curve, $S_t = \mathbb{E}_C\bigl[\frac{1}{HWC_{\mathrm{ch}}}\sum_i \mathbf{1}[c_t^{(i)} = c_0^{(i)}]\bigr]$, the expected fraction of pixels not yet absorbed at step $t$, by Monte Carlo on training images, and set $t(s)$ to the step with $S_t = s$, so the prior sees the kind of input it was trained on at that step. $S_t$ falls monotonically from 1 to 0, so each budget has one timestep; we estimate it on 20 training batches at every tenth step and interpolate (on MNIST, $s = 0.1$ gives $t = 935$ of $T = 1000$). The timestep follows the target budget, not the density a mask actually achieves: a mask far below budget presents many more mask tokens than the prior expects at $t(s)$, so it is queried off its training distribution. The surrogate is therefore only meaningful when the budget holds.

\subsection{Mask generator and objective}
The mask generator concatenates the $8 \times 8$ summary in $Z$ with an MLP embedding of $s$ (and a class embedding on MNIST), each broadcast over the $8 \times 8$ grid, and decodes per-pixel keep/drop logits with a convolutional encoder--decoder (Figure~\ref{fig:pipeline}). Each pixel is kept independently with probability $p_i = \sigma(a_i)$, $a_i$ the difference of its two logits, and the generator is trained with DisARM \citep{dong2020disarm}, an unbiased gradient estimator for binary masks; Section~\ref{sec:images} explains why we do not use straight-through Gumbel-Softmax. The loss is
\begin{equation}
\mathcal{L} = \mathcal{H}_{\mathrm{all}} + \lambda\,\mathbb{E}\bigl[(\bar{X} - s)^2\bigr],
\end{equation}
where $\bar{X}$ is the fraction of pixels a sampled mask keeps and $s$ is drawn per sample from the budget range (Table~\ref{tab:hyper}). The first term is the whole-image surrogate computed on that mask's observation at $t(s)$; the second, weighted by $\lambda$, holds the expected density at the budget. The D3PM is frozen: its weights never change, it only scores the observations of sampled masks, and no gradient passes through it, since DisARM estimates the gradient with respect to the keep logits from the loss values of antithetic pairs of masks.

\subsection{Sequential greedy information gain}
\label{sec:closed-method}
Measuring pixel $i$ noiselessly reveals $C_i$, so its information gain given the current observation $y$ is
\begin{equation}
I(C; Y_i \mid y) = H(Y_i \mid y) - H(Y_i \mid C, y) = H(C_i \mid y),
\end{equation}
the prior's entropy at $i$. Greedy acquisition therefore needs no surrogate: at each acquisition step it measures the $k_r$ unobserved pixels of highest $H(C_i \mid y)$, then recomputes the posterior. Taking all $k$ pixels in one step is suboptimal \citep{kirsch2019batchbald}; splitting the budget over $R$ steps lets each step see what the last one revealed. The first step starts either from a small random probe or from no observations at all, in which case the label in the side information is the only thing placing the first pixels. For this method to work the prior must distinguish ``observed black'' from ``unobserved'', which the separate mask state of Section~\ref{sec:formulation} provides.

\section{One-shot masks on MNIST and CIFAR-10}
\label{sec:images}

\subsection{Setup}
We compare five methods here and in Section~\ref{sec:closed}: \textbf{random} (budget-exact uniform masks), \textbf{variable density} (density decaying away from the image center), \textbf{LOUPE} \citep{bahadir2020loupe} (a learned probabilistic mask, ported to the pixel domain and trained per budget with its own U-Net), \textbf{one-shot (ours)} (the generator trained on $\mathcal{H}_{\mathrm{all}}$ with DisARM) and \textbf{sequential (ours)} (greedy information gain over 16 steps), all reconstructed by the same prior unless stated otherwise. Both datasets use the encoding of Section~\ref{sec:formulation}. MNIST is padded to $32 \times 32$ and binarized ($N{=}3$ states); CIFAR-10 is $32 \times 32 \times 3$ with $8$ intensity bins and the mask token ($N{=}9$). The frozen D3PMs use $T{=}1000$ steps and a cosine schedule: a small convolutional $c_0$-predictor on MNIST and a DiT ($d{=}1024$) on CIFAR-10. Mask generators are trained with $\mathcal{H}_{\mathrm{all}}$ for 200 epochs. All numbers are on the hold-out test split ($2560$ images), over 5 seeds. Hidden pixels are reconstructed at the D3PM's argmax; we report errors per image and foreground recovery (digit pixels recovered) on MNIST, and PSNR with hidden pixels at the posterior mean on CIFAR-10. Learned masks are drawn on budget: the $\mathrm{round}(sHW)$ pixels with the largest keep log-odds plus logistic noise. Appendix~\ref{app:details} gives the hyperparameters and the full evaluation protocol.

\paragraph{Gradient estimator.}
The straight-through Gumbel-Softmax gradient does not train these generators. Against the exact gradient of the expected loss, estimated by flipping one pixel of a sampled mask at a time, it has cosine $-0.73$ to $-0.97$ on the generator's weights, so it points uphill, and no checkpoint trained with it reached a lower objective than a random mask. DisARM \citep{dong2020disarm} has cosine $0.94$ to $0.99$ at the same cost, and under it the objective falls from the first epoch (Appendix~\ref{app:details}).

\paragraph{Results.}
\begin{table}[!htbp]
\centering
\caption{What the one-shot generator gets from its side information: the target's own $8 \times 8$ summary, another test image's summary, or none (an all-zero summary). Budget-exact sampled masks, hold-out test split, 5 seeds; seed spread at most $0.10$ errors per image and $0.02$~dB. On MNIST the generator outperforms random even without the true summary; on CIFAR-10 its gain needs it.}
\label{tab:side-info}
\small
\setlength{\tabcolsep}{4pt}
\begin{tabular}{lccccc}
\hline
 & \multicolumn{2}{c}{\textbf{MNIST} (errors/img)} & \multicolumn{3}{c}{\textbf{CIFAR-10} (PSNR, dB)} \\
\textbf{Summary} & $s = 0.1$ & $0.3$ & $s = 0.1$ & $0.3$ & $0.5$ \\
\hline
target's own & 23.0 & 4.74 & 20.45 & 24.80 & 28.01 \\
another image's & 26.8 & 5.22 & 20.12 & 24.25 & 27.40 \\
none & 26.4 & 5.31 & 19.89 & 23.78 & 26.42 \\
random (no generator) & 46.7 & 24.0 & 20.14 & 24.33 & 27.47 \\
\hline
\end{tabular}
\end{table}

On MNIST the generator lowers $\mathcal{H}_{\mathrm{all}}$ below random at every budget, and the gain carries over to criteria it was not trained on: at $s = 0.1$ it recovers $90\%$ of digit pixels against $78\%$ and makes $23.0$ errors per image against $46.7$, and $41\%$ of its observed pixels are digit pixels, against $9\%$ for a random mask (Figure~\ref{fig:examples-mnist}; every budget and metric in Appendix~\ref{app:oneshot}).

\begin{figure}[!htbp]
\centering
\includegraphics[width=\textwidth]{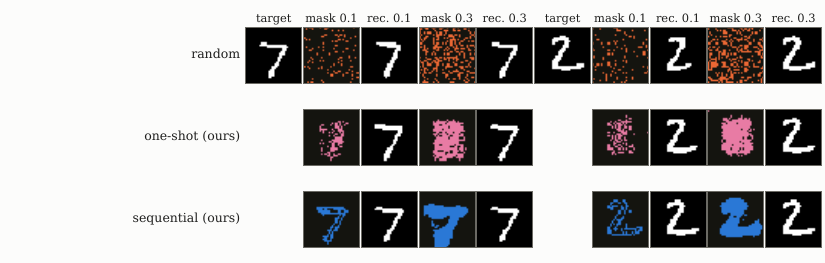}
\caption{MNIST test images at $s = 0.1$ and $0.3$, one row per method; masks are tinted by method, and each reconstruction keeps the observed pixels and fills the rest from the prior. The one-shot generator, trained with DisARM and sampled on budget, covers the region where the digit lies, densely by $s = 0.3$; sequential acquisition (label-greedy, 16 steps; Section~\ref{sec:closed}) places its pixels on and around the stroke, because the prior's entropy concentrates there, as Figure~\ref{fig:pipeline}(b) shows step by step. On all $2560$ test images the one-shot generator makes $22.9$ errors per image at $s = 0.1$ against $46.6$ for random ($4.7$ against $24.0$ at $s = 0.3$), and sequential acquisition makes $5.6$ and $0.01$.}
\label{fig:examples-mnist}
\end{figure}

The generator sees an $8 \times 8$ summary of the target image (Table~\ref{tab:side-info}). Replacing it with another test image's summary raises errors at $s = 0.1$ from $23.0$ to $26.8$, and removing it gives $26.4$, still far below random ($46.7$). With deterministic top-$k$ selection, the mode of the learned masks, the summary matters more ($27.6$, $39.7$ and $42.1$ against $46.5$), and by $s = 0.3$ it hardly matters ($3.66$ to $4.54$ errors against $24.0$). Sequential acquisition, which uses only the label, still makes about $4\times$ fewer errors at $s = 0.1$ ($5.6$; Section~\ref{sec:closed}).

On CIFAR-10 the gain is small, $0.3$ to $0.7$~dB PSNR, against a seed spread of at most $0.02$~dB. The whole gain comes from the target-derived summary (Table~\ref{tab:side-info}): with another image's summary the generator is level with random, and with none it is worse. At $s = 0.1$ its objective ends slightly above random's while its PSNR keeps improving, and a plain Bernoulli draw overshoots the budget (density $0.187$) and appears to gain $2.6$~dB, most of it from the overshoot (Appendix~\ref{app:oneshot}).

\section{Sequential acquisition}
\label{sec:closed}

\begin{figure}[!t]
\centering
\includegraphics[width=0.92\textwidth]{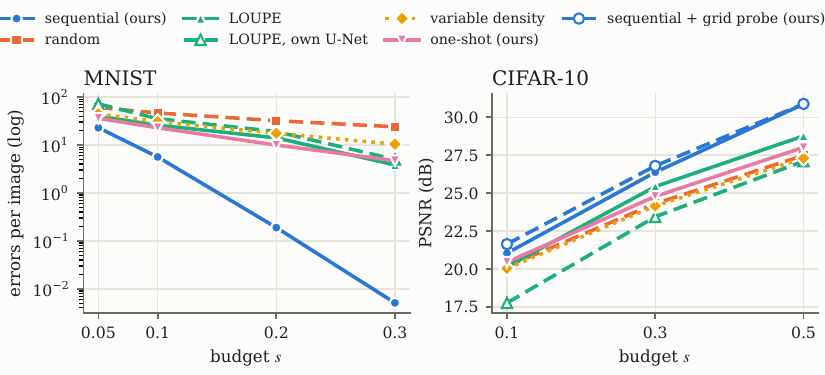}
\caption{Sequential acquisition against the baselines on the test split; every mask is reconstructed by the same prior except ``LOUPE, own U-Net''. MNIST: label-greedy with 16 steps, errors per image (log scale). CIFAR-10: variance-greedy with 16 steps, alone and after a jittered-grid probe, PSNR. The grid probe is used on CIFAR-10 only: on MNIST, starting from the label is better than starting from a probe (Table \ref{tab:closed}). LOUPE is shown at the better of its top-$k$ and budget-exact sampled masks (Appendix~\ref{app:sequential}). The one-shot mask sees the target's own $8 \times 8$ summary.}
\label{fig:baselines}
\end{figure}

We apply Section~\ref{sec:closed-method} to the MNIST prior on the same $2560$ test images. Strategies share images, probes and random masks, so comparisons are paired, and the probe takes $20\%$ of the budget. Because recovery saturates once whole digits are reconstructed, we report errors per image and the fraction of images reconstructed exactly (Table~\ref{tab:closed}).

\begin{table}[!htbp]
\centering
\caption{MNIST errors per test image (lower is better; fraction of images reconstructed exactly in parentheses).}
\label{tab:closed}
\small
\setlength{\tabcolsep}{4pt}
\begin{tabular}{lcccc}
\hline
\textbf{Strategy} & $s = 0.05$ & $s = 0.1$ & $s = 0.2$ & $s = 0.3$ \\
\hline
random & 60.5 (0\%) & 46.7 (0\%) & 32.1 (0\%) & 24.0 (0\%) \\
probe + greedy, 1 step & 40.0 (0\%) & 26.1 (1\%) & 9.22 (14\%) & 1.97 (50\%) \\
probe + greedy, 16 steps & 26.9 (4\%) & 8.45 (17\%) & 0.42 (83\%) & 0.02 (99\%) \\
label-greedy, 4 steps & 26.8 (4\%) & 10.0 (14\%) & 0.64 (81\%) & 0.03 (99\%) \\
\textbf{label-greedy, 16 steps} & \textbf{22.9} (6\%) & \textbf{5.63} (26\%) & \textbf{0.19} (92\%) & \textbf{0.01} (100\%) \\
\hline
\end{tabular}
\end{table}

A single greedy step after a random probe already removes $44\%$ of the errors a random mask leaves at $s = 0.1$, and $92\%$ at $s = 0.3$. The best method starts from no observations and takes $16$ steps: at $s = 0.1$ it leaves $5.6$ errors per image against $46.7$ for random, and at $s = 0.2$ it reconstructs $92\%$ of test digits exactly, against none for random. Starting from the label is better than starting from a random probe at every budget: the class-conditional prior already knows where a digit of that class is uncertain. On CIFAR-10 the opposite holds: label-greedy with $16$ steps is level with a random mask at $s = 0.1$ and $0.3$ and about $2$~dB behind a random probe followed by variance-greedy steps, since a CIFAR-10 class label says little about where an image's informative pixels are. Against the baselines (Figure~\ref{fig:baselines}), sequential acquisition outperforms LOUPE, variable density and the one-shot mask at every budget on both datasets; on CIFAR-10 it reaches $21.6$, $26.8$ and $30.9$~dB at $s = 0.1$, $0.3$ and $0.5$, $1.2$ to $2.9$~dB above the one-shot mask. Appendix~\ref{app:sequential} gives the per-step curves and example masks.

\section{FastMRI}
\label{sec:mri}

\subsection{Setup}
We use the fastMRI single-coil knee data \citep{zbontar2018fastmri}, one mid-slice per volume: $973$ training and $199$ hold-out validation slices at $300 \times 300$ pixels. A mask selects rows of k-space around a $32$-row auto-calibration (ACS) block that is always sampled, and the budget is enforced exactly by top-$k$. Side information $Z$ is a central scout reconstruction, and the one-shot policy is an instance-adaptive row selector trained on zero-filled reconstruction error plus the coarse row-profile entropy surrogate of a frozen D3PM. Two U-Nets are trained on the training split under heuristic masks only: \textbf{U-Net A} ($1.9$M parameters), which masks are optimized against, and \textbf{U-Net B} ($4.3$M parameters, different seed and mask draws), a hold-out judge used only for scoring. We report per-slice NMSE, $\lVert \hat{c} - c \rVert^2 / \lVert c \rVert^2$, under U-Net B; stochastic baselines are averaged over 5 seeds, and a margin inside two standard deviations is not counted as an improvement.

\subsection{Results}
\textbf{The entropy surrogate disagrees with reconstruction quality.} At $s = 0.25$ the learned policy attains the lowest entropy of any method and still loses on NMSE to every heuristic tested. Table~\ref{tab:recon-gain} explains why: masks that spread out gain most from a reconstructor, while center-concentrated masks gain least, because a reconstructor with a prior already predicts the low-frequency center: spread-out masks (ACS + equispaced or random) gain about $30\%$, while the energy oracle and the learned policies gain only $14$--$16\%$.

\begin{table}[!htbp]
\centering
\caption{Val NMSE at $s = 0.25$. Gain is the relative NMSE reduction from zero-filled to U-Net B reconstruction, $1 - \mathrm{NMSE}_{\mathrm{B}}/\mathrm{NMSE}_{\mathrm{zf}}$. Center-concentrated masks gain least from a reconstructor that carries a prior.}
\label{tab:recon-gain}
\small
\begin{tabular}{lccc}
\hline
\textbf{Mask} & \textbf{Zero-filled} & \textbf{U-Net B} & \textbf{Gain} \\
\hline
ACS + equispaced & 0.03820 & 0.02623 & 31.3\% \\
ACS + random & 0.03819 & 0.02696 & 29.4\% \\
static search (ours) & 0.03283 & 0.02389 & 27.2\% \\
ACS + variable density & 0.03382 & 0.02474 & 26.8\% \\
learned policy, static (ours) & 0.03431 & 0.02887 & 15.9\% \\
learned policy, adaptive (ours) & 0.03337 & 0.02840 & 14.9\% \\
energy oracle & 0.03147 & 0.02715 & 13.7\% \\
\hline
\end{tabular}
\end{table}

\begin{table}[!htbp]
\centering
\caption{Hold-out val NMSE under the judge U-Net B. ACS + VD: mean $\pm$ sd over 5 seeds. Sequential: rows ranked by the coarse row-profile prior's entropy, 16 steps.}
\label{tab:mri-main}
\small
\setlength{\tabcolsep}{4pt}
\begin{tabular}{lcccc}
\hline
$s$ & \textbf{Static search (ours)} & \textbf{ACS + VD} & \textbf{LOUPE} & \textbf{Sequential} \\
\hline
0.25 & \textbf{0.02389} & 0.02471 $\pm$ 0.00006 & 0.02759 & 0.02589 \\
0.40 & \textbf{0.01860} & 0.01889 $\pm$ 0.00005 & 0.02213 & 0.01992 \\
0.50 & \textbf{0.01523} & 0.01546 $\pm$ 0.00007 & 0.01995 & 0.01559 \\
0.75 & \textbf{0.00751} & 0.00764 $\pm$ 0.00001 & 0.00966 & 0.00807 \\
\hline
\end{tabular}
\end{table}

\begin{figure}[!htbp]
\centering
\includegraphics[width=0.9\textwidth]{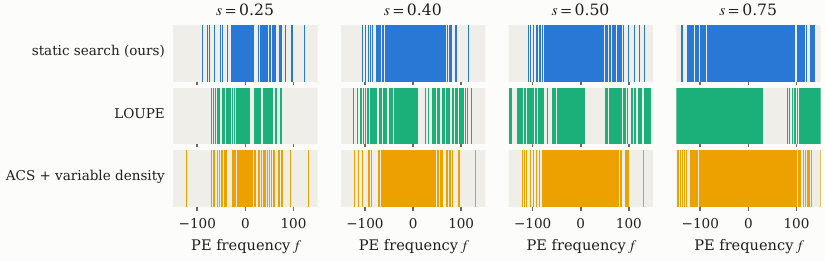}
\caption{Phase-encode rows sampled by the static search mask, LOUPE and ACS + variable density ($f = 0$ is DC); Figure~\ref{fig:mri-rows-all} in Appendix~\ref{app:mri} shows every static mask.}
\label{fig:mri-rows}
\end{figure}

\begin{figure}[!tb]
\centering
\includegraphics[width=0.85\textwidth]{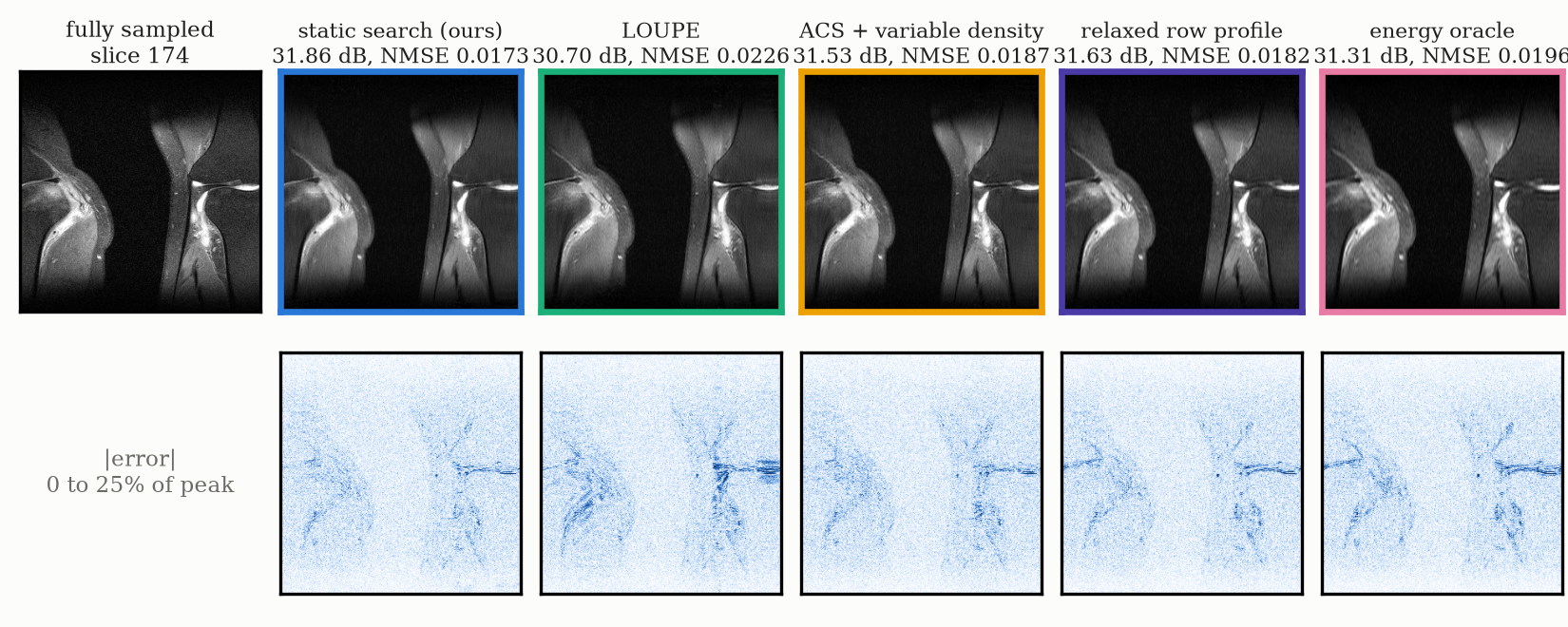}
\caption{A validation slice of median difficulty for ACS + variable density, at $s = 0.25$, reconstructed by the hold-out judge. Top: reconstructions. Bottom: absolute error $|$reconstruction $-$ fully sampled$|$, from 0 (white) to a quarter of the slice's 99.5th-percentile intensity (dark blue). PSNR and NMSE are per slice.}
\label{fig:mri-examples}
\end{figure}

\begin{figure}[!t]
\centering
\includegraphics[width=\textwidth]{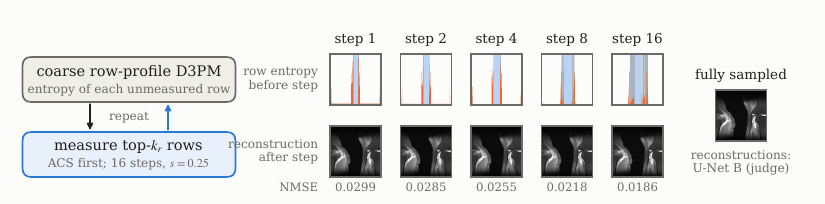}
\caption{Sequential acquisition on the slice of Figure~\ref{fig:mri-examples} at $s = 0.25$: from the ACS block, each of 16 steps adds the unmeasured k-space rows of highest prior entropy. Top: per-row entropy before the step (DC in the middle; blue: measured rows). Bottom: U-Net B reconstruction after the step, with its NMSE.}
\label{fig:mri-sequential}
\end{figure}

\textbf{Optimizing the discrete mask against the reconstructor works better.} A greedy row-swap search on the hard mask, in which the gradient proposes candidate swaps and exact evaluation on the training split accepts them, is best at every budget (Table~\ref{tab:mri-main}), also on PSNR and under zero-filled reconstruction; a little over half of its gain comes from starting at the best fixed variable-density draw, the rest from the swaps. LOUPE, restricted to whole phase-encode rows, is worse than every heuristic (Figure~\ref{fig:mri-rows} shows the rows each mask samples), and a relaxed row profile trained through a straight-through top-$k$ mask improves on variable density only at $s = 0.50$. Sequential acquisition with the only prior that trained well here, a coarse row-profile D3PM, loses to variable density at $s = 0.25$ and $0.40$ and never reaches the search mask's NMSE: after the ACS block the coarse profile carries little slice-specific information. Figure~\ref{fig:mri-examples} shows a validation slice of median difficulty at $s = 0.25$: the search mask has the lowest NMSE ($0.0173$, against $0.0187$ for variable density), and LOUPE's errors are largest along tissue edges. Figure~\ref{fig:mri-sequential} follows sequential acquisition on the same slice: its 16 steps lower NMSE from $0.0299$ to $0.0186$, level with variable density and behind the search mask. Appendix~\ref{app:mri} gives the remaining figures, every sequential variant and the LOUPE details.

\section{Discussion and Conclusion}
\label{sec:conclusion}

We cast budgeted acquisition with side information as maximizing $I(Y;C)$ and used a frozen absorbing D3PM in two ways: as a surrogate for $H(C \mid Y)$ that a one-shot mask generator is trained against, and as the posterior for sequential greedy acquisition. The two behave differently. For noiseless pixels, the criterion of sequential acquisition is the information gain itself: in 16 steps it leaves $8\times$ fewer errors than random on MNIST at $s = 0.1$ and gains $0.9$--$3.4$~dB on CIFAR-10. The one-shot generator gains less, and only with care: it needs an unbiased gradient estimator, whole-image averaging and budget-exact sampling, and on CIFAR-10 its whole gain comes from the target's own summary. The prior's encoding matters too: a separate mask state gives $10$--$15\%$ fewer errors than a prior in which the mask shares its value with black (Appendix~\ref{app:token}).

\paragraph{Limitations.} The surrogate averages per-pixel entropies, an upper bound that ignores dependence between pixels, so greedy steps can pick redundant pixels \citep{kirsch2019batchbald}; the categorical entropy also ignores the order of intensity bins, one reason the objective and PSNR can part ways on CIFAR-10. The information-gain identity assumes noiseless measurements. Our images are small ($32 \times 32$), and sequential acquisition costs one pass of the prior per step, a one-shot mask one pass of the generator. On fastMRI the coarse row-profile prior carries little slice-specific information, so sequential acquisition trails a static mask optimized against a reconstructor.

\paragraph{Future work.} Stronger k-space priors, batch-aware criteria, noisy measurements, and an amortized sequential policy with the speed of a one-shot mask, which would suit real-time sensing such as adaptive LiDAR in autonomous driving \citep{bergman2020deep}.

\subsection*{AI use statement}
We used Claude Code (Anthropic) to assist with quick model architecture experimentation, code debugging, and monitoring training jobs; Claude was also used to check grammatical errors, detect ambiguous or confusing phrasing in some paragraphs, and help with exploring alternative designs; Gemini and Claude were also used to select pleasant color palettes for plotting, and help with formatting LaTeX figures correctly to avoid overflow. LLM-based Google Scholar Labs was used to identify relevant references. The research question, problem formulation, core implementation, and main writing are the authors'; no LLM was used during the ideation and proof-of-concept phase. The authors reviewed all code, results and text and take full responsibility for the content of this paper.

\subsection*{Reproducibility statement}
All datasets are public: MNIST, CIFAR-10 and the fastMRI single-coil knee data \citep{zbontar2018fastmri}. Sections~\ref{sec:method} and~\ref{sec:images} and Table~\ref{tab:hyper} give the priors, the mask-generator architecture and its training settings; Sections~\ref{sec:closed-method} and~\ref{sec:closed} the sequential acquisition procedure; Section~\ref{sec:mri} the reconstructors, masks and splits used on fastMRI; and Appendix~\ref{app:token} the comparison of priors. Every image result is on the hold-out test split with fixed evaluation seeds, and we report the spread over seeds. We will release the code, trained checkpoints and evaluation scripts after the review period.

\bibliographystyle{iclr2027_conference}
\bibliography{references}

\appendix
\renewcommand{\thesubsection}{\Alph{subsection}}
\section*{Appendix}

\subsection{Experimental details}
\label{app:details}

\paragraph{Priors.} Each D3PM is initialized from a D3PM trained on the same data without the mask state and then trained under the encoding of Section~\ref{sec:formulation} for $60$ epochs on MNIST and $30$ on CIFAR-10. Table~\ref{tab:hyper} lists the mask-generator settings.

\begin{table}[!htbp]
\centering
\caption{Mask-generator training; DisARM uses one antithetic pair per image.}
\label{tab:hyper}
\small
\begin{tabular}{lcc}
\hline
\textbf{Setting} & \textbf{MNIST} & \textbf{CIFAR-10} \\
\hline
Batch size & 512 & 128 \\
Learning rate & $3 \times 10^{-4}$ & $2 \times 10^{-5}$ \\
Sparsity weight $\lambda$ & 10 & 10 \\
Budget range & $[0.05, 0.95]$ & $[0.1, 0.9]$ \\
Gradient estimator & DisARM & DisARM \\
Epochs & 200 & 200 \\
\hline
\end{tabular}
\end{table}

\paragraph{Gradient estimator.} For independent per-pixel masks the exact gradient of the expected loss is $p_i(1-p_i)\,\mathbb{E}\bigl[\mathcal{L}(X_i{=}1) - \mathcal{L}(X_i{=}0)\bigr]$, which we estimate by flipping one pixel of a sampled mask at a time. On the MNIST generator, untrained and after training, the straight-through gradient has cosine $-0.73$ to $-0.97$ with it on the generator's weights. Trained with it, the generator's whole-image entropy rose at learning rate $3 \times 10^{-4}$ and did not fall at $10^{-4}$ or $3 \times 10^{-5}$, and no checkpoint reached a lower objective than a random mask. DisARM needs two forward passes of the frozen prior per antithetic pair and no backward pass through it, and has cosine $0.94$ to $0.99$ with the exact gradient.

\paragraph{Evaluation.} All image numbers are on the hold-out test split ($20$ batches of $128$), at fixed budgets $s \in \{0.1, 0.3, 0.5, 0.7\}$, against budget-exact uniformly random masks. Besides the surrogate objective itself, we report several criteria the mask was not optimized for. We reconstruct each image with its observed pixels at their measured values and its hidden pixels at the D3PM's argmax, and score it by full-image accuracy and by foreground recovery, the fraction of nonzero (digit) pixels recovered correctly; since ${\sim}90\%$ of MNIST pixels are background, an all-black guess already reaches ${\sim}0.90$ full-image accuracy, so foreground recovery is the informative one. On CIFAR-10 we add PSNR, with hidden pixels at the D3PM's posterior mean on a $[0,1]$ scale; the reference is the $8$-bin quantized target the D3PM models, not the original image, so quantization error is excluded from every PSNR we report. The informative fraction is the share of observed pixels that are not black (on MNIST, digit pixels). We also report accuracy on the hidden pixels, overall and on the hidden foreground; these are biased against a mask that observes more of the digit, so we read them alongside the whole-reconstruction scores. Learned masks are drawn on budget: the $\mathrm{round}(sHW)$ pixels with the largest keep log-odds plus logistic noise, which is a Bernoulli draw's own noise with its threshold moved to meet the budget exactly. A plain draw overshoots it, by up to $87\%$ on CIFAR-10.

\subsection{Additional one-shot results}
\label{app:oneshot}

\begin{figure}[!htbp]
\centering
\includegraphics[width=\textwidth]{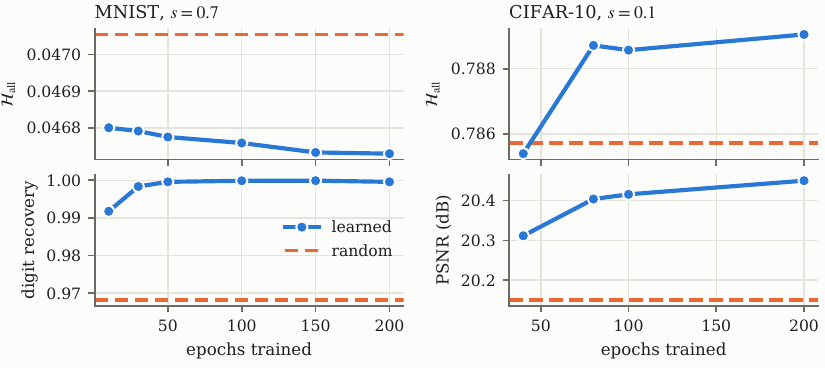}
\caption{Top: the value of the whole-image surrogate objective the mask is trained on, with DisARM. Bottom: a criterion the mask is not optimized for, on the same checkpoints, with budget-exact sampled masks. On MNIST at $s = 0.7$ both improve together, and digit recovery stays above random throughout ($0.992$ at epoch 10 and $1.000$ at 200, against $0.968$). On CIFAR-10 at $s = 0.1$ they part ways: the objective rises above random's after epoch 40 while PSNR keeps improving and stays above random ($20.31$~dB at epoch 40, $20.45$ at 200, against $20.15$).}
\label{fig:divergence}
\end{figure}

\begin{figure}[!htbp]
\centering
\includegraphics[width=\textwidth]{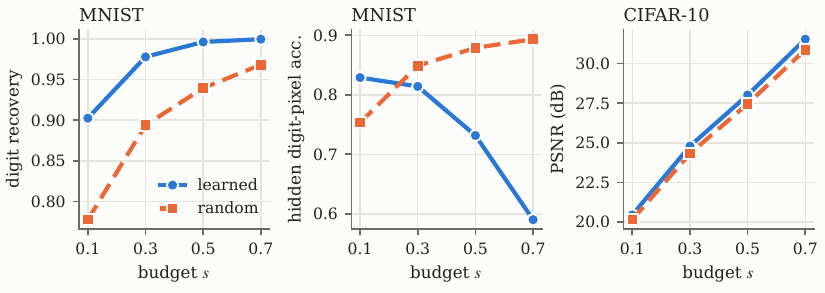}
\caption{The MNIST advantage in digit recovery (left) holds at every budget and narrows as the budget grows ($0.903$ vs.\ $0.777$ at $s = 0.1$, $1.000$ vs.\ $0.968$ at $s = 0.7$); accuracy on the digit pixels the mask leaves hidden (centre) is worse than random for every $s \ge 0.3$, a measure biased against a mask that observes more of the digit. Mask trained with DisARM, final checkpoint, budget-exact sampled masks. On CIFAR-10 (right) the mask is $0.3$ to $0.7$~dB above random at every budget.} 
\label{fig:metrics-vs-budget}
\end{figure}

\begin{figure}[!htbp]
\centering
\includegraphics[width=\textwidth]{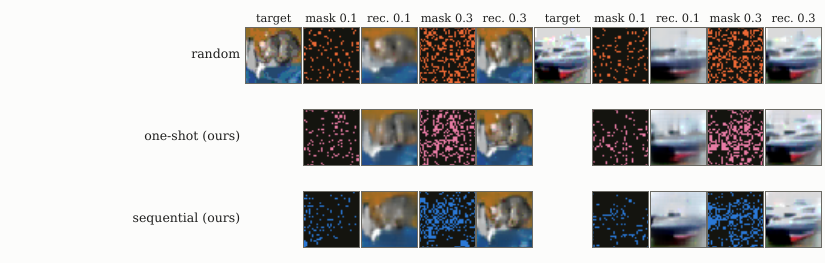}
\caption{CIFAR-10 test images at $s = 0.1$ and $0.3$, one row per method; masks are tinted by method, and each reconstruction keeps the observed pixels and fills the rest with the prior's posterior mean. The one-shot generator, trained with DisARM and sampled on budget, scatters its pixels over the whole image much as a random mask does; sequential acquisition (variance-greedy, 16 steps; Section~\ref{sec:closed}) spreads them more evenly and more densely where the image has structure.}
\label{fig:examples-cifar}
\end{figure}

Tables~\ref{tab:mnist-open} and~\ref{tab:cifar-open} give every metric for the one-shot generators. On MNIST the objective and the criterion improve together over training (Figure~\ref{fig:divergence}): at $s = 0.7$, digit recovery stays above random from epoch 10 ($0.992$) to epoch 200 ($1.000$, against $0.968$). The advantage narrows as the budget grows (Figure~\ref{fig:metrics-vs-budget}); only accuracy on the digit pixels the mask leaves hidden falls below random, from $s = 0.3$, a measure biased against a mask that observes more of the digit. On CIFAR-10 the generator lowers the surrogate below random only from $s = 0.3$; at $s = 0.1$ its objective ends slightly above random's while its PSNR keeps improving (Figure~\ref{fig:divergence}). Figure~\ref{fig:examples-cifar} shows example masks on CIFAR-10.

\begin{table}[!htbp]
\centering
\caption{MNIST mask generator trained with $\mathcal{H}_{\mathrm{all}}$, learned / random, hold-out test split, 5 seeds, budget-exact sampled masks (bold = better). Hidden-fg acc.\ scores only the digit pixels left hidden; it is biased against good masks, since a mask that observes most of the digit leaves only the pixels the prior finds hardest.}
\label{tab:mnist-open}
\small
\setlength{\tabcolsep}{2.5pt}
\begin{tabular}{ccccccc}
\hline
$s$ & $\mathcal{H}_{\mathrm{all}}$ & Full acc. & Fg recovery & Informative & Hidden-fg acc. & Errors/img \\
\hline
0.1 & \textbf{0.0564} / 0.0685 & \textbf{0.978} / 0.954 & \textbf{0.903} / 0.777 & \textbf{0.41} / 0.09 & \textbf{0.829} / 0.753 & \textbf{23.0} / 46.7 \\
0.3 & \textbf{0.0495} / 0.0527 & \textbf{0.995} / 0.977 & \textbf{0.978} / 0.894 & \textbf{0.28} / 0.09 & 0.813 / \textbf{0.849} & \textbf{4.74} / 24.0 \\
0.5 & \textbf{0.0476} / 0.0488 & \textbf{0.999} / 0.986 & \textbf{0.996} / 0.939 & \textbf{0.19} / 0.09 & 0.734 / \textbf{0.879} & \textbf{0.68} / 14.4 \\
0.7 & \textbf{0.0467} / 0.0471 & \textbf{1.000} / 0.992 & \textbf{1.000} / 0.968 & \textbf{0.13} / 0.09 & 0.578 / \textbf{0.893} & \textbf{0.06} / 7.88 \\
\hline
\end{tabular}
\end{table}

\begin{table}[!htbp]
\centering
\caption{CIFAR-10 mask generator trained with $\mathcal{H}_{\mathrm{all}}$, learned / random, hold-out test split, 5 seeds, budget-exact sampled masks (bold = better). Non-black rec.\ is the fraction of non-black pixels reconstructed correctly.}
\label{tab:cifar-open}
\small
\setlength{\tabcolsep}{3.5pt}
\begin{tabular}{ccccc}
\hline
$s$ & $\mathcal{H}_{\mathrm{all}}$ & Full acc. & Non-black rec. & PSNR (dB) \\
\hline
0.1 & 0.790 / \textbf{0.786} & \textbf{0.675} / 0.670 & \textbf{0.676} / 0.670 & \textbf{20.45} / 20.14 \\
0.3 & \textbf{0.420} / 0.435 & \textbf{0.824} / 0.816 & \textbf{0.824} / 0.815 & \textbf{24.80} / 24.33 \\
0.5 & \textbf{0.236} / 0.250 & \textbf{0.901} / 0.894 & \textbf{0.901} / 0.893 & \textbf{28.01} / 27.47 \\
0.7 & \textbf{0.116} / 0.126 & \textbf{0.951} / 0.946 & \textbf{0.951} / 0.946 & \textbf{31.54} / 30.87 \\
\hline
\end{tabular}
\end{table}

\subsection{Additional sequential results}
\label{app:sequential}

\begin{figure}[!htbp]
\centering
\includegraphics[width=\textwidth]{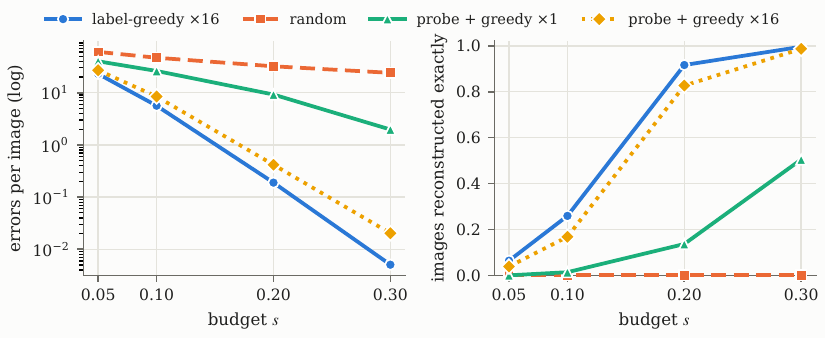}
\caption{Sequential acquisition on MNIST (exact values in Table~\ref{tab:closed}). Left: errors per image, log scale. Right: fraction of test images reconstructed exactly. }
\label{fig:closed}
\end{figure}

\begin{figure}[!htbp]
\centering
\includegraphics[width=\textwidth]{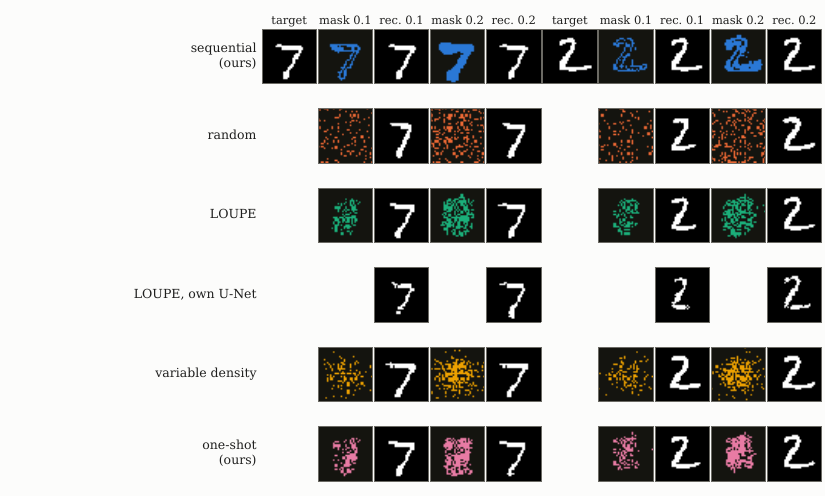}
\caption{MNIST test images at $s = 0.1$ and $0.2$: each method's mask, tinted in its colour, and the reconstruction under the prior (LOUPE's own U-Net in its own row). LOUPE's masks are sampled from its learned map.}
\label{fig:baselines-mnist}
\end{figure}

\begin{figure}[!htbp]
\centering
\includegraphics[width=\textwidth]{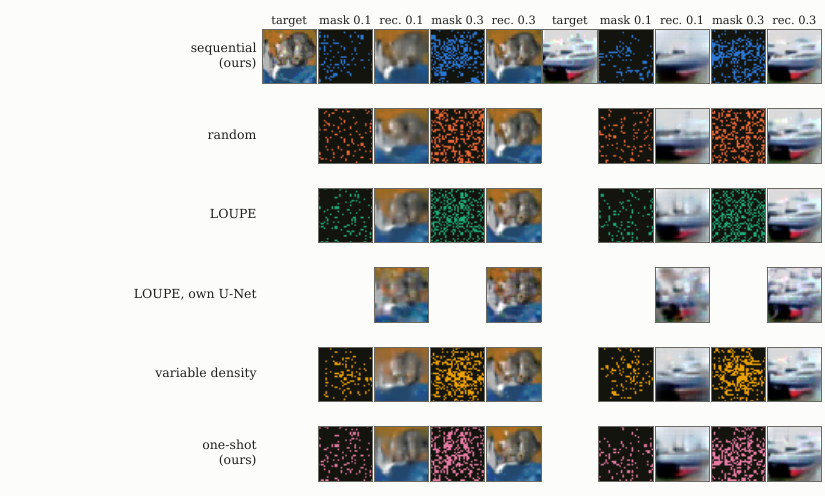}
\caption{CIFAR-10 test images at $s = 0.1$ and $0.3$, as in Figure~\ref{fig:baselines-mnist}; the sequential-acquisition row is variance-greedy with 16 steps. The one-shot mask, trained with DisARM and sampled on budget, scatters its pixels over the image much as a random mask does.}
\label{fig:baselines-cifar}
\end{figure}

Figure~\ref{fig:closed} plots Table~\ref{tab:closed}. Going from $1$ to $16$ steps reduces errors $3\times$ at $s = 0.1$ and $22\times$ at $s = 0.2$. On CIFAR-10, label-greedy with $16$ steps reaches $20.0$ and $24.4$~dB at $s = 0.1$ and $0.3$, against $20.2$ and $24.3$ for random and $21.1$ and $26.4$ for a random probe followed by variance-greedy steps; with $4$ steps it falls below $15$~dB. In Figure~\ref{fig:baselines}, LOUPE is shown at the better of two test protocols at each budget: the top-$k$ of its learned map, as for MRI, or a budget-exact sample from it, the distribution it trained on (top-$k$ degenerates to the image border on CIFAR-10 at $s = 0.1$, where the map never polarized). Figures~\ref{fig:baselines-mnist} and~\ref{fig:baselines-cifar} show example masks of every method.

\subsection{fastMRI details}
\label{app:mri}

\paragraph{Reconstructors.} Both U-Nets are trained under six heuristic mask families, random, equispaced and variable density, each with and without the ACS block, with $s \sim \mathcal{U}(0.10, 0.75)$.

\begin{figure}[!htbp]
\centering
\includegraphics[width=\textwidth]{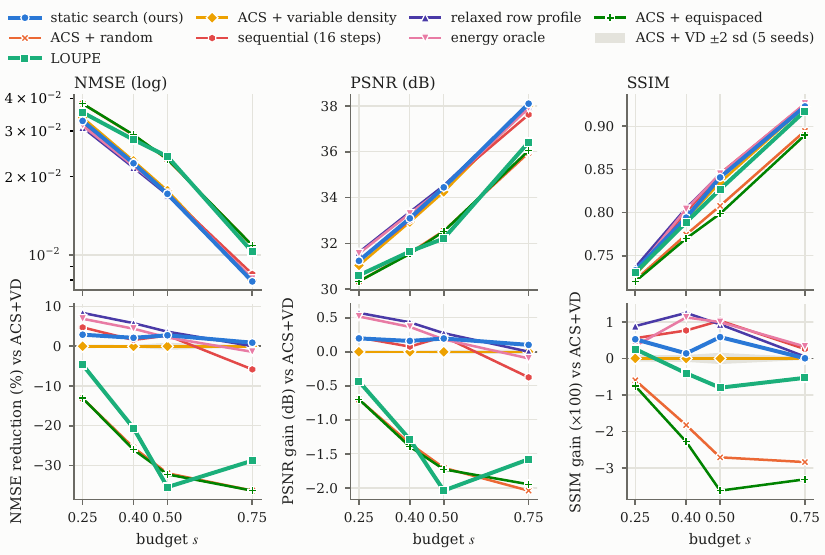}
\caption{The same masks under zero-filled reconstruction, with no network. The sequential-acquisition line has lower NMSE than variable density here at $s \le 0.50$.}
\label{fig:mri-zf}
\end{figure}

\begin{figure}[!htbp]
\centering
\includegraphics[width=\textwidth]{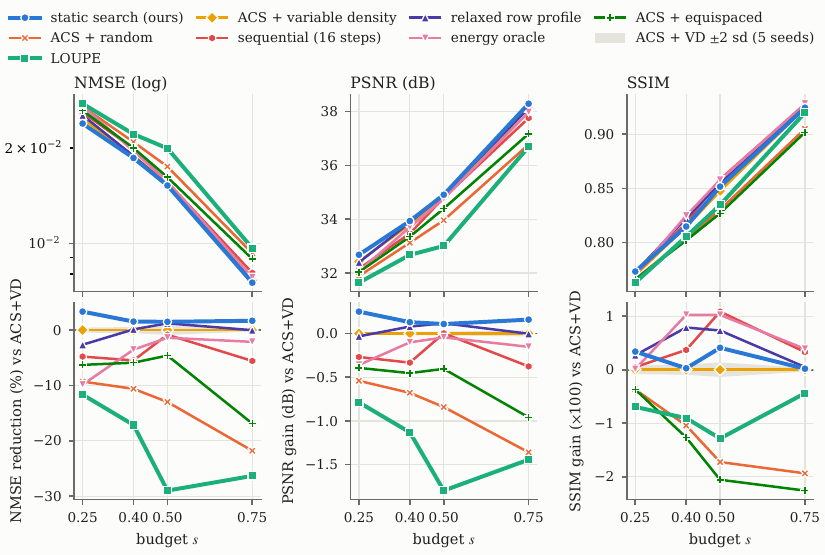}
\caption{Row masks on the 199 hold-out validation slices, reconstructed by the hold-out judge U-Net B. Top: NMSE (log scale), PSNR and SSIM against the budget. Bottom: gain over ACS + variable density. The static search mask is best on NMSE and PSNR at every budget and LOUPE is last; the center-concentrated relaxed row profile and energy oracle, with sequential acquisition, lead on SSIM at $s = 0.40$--$0.75$. The sequential-acquisition line (16 steps; Table~\ref{tab:mri-closed}) trails variable density on NMSE at every budget except $s = 0.50$, where it is level.}
\label{fig:mri-judge}
\end{figure}

\begin{figure}[!htbp]
\centering
\includegraphics[width=\textwidth]{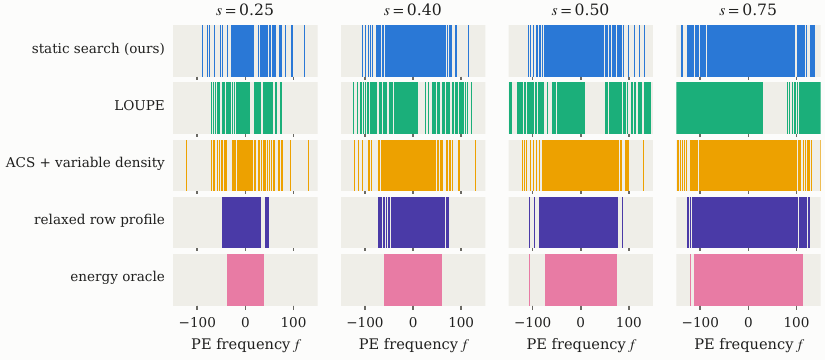}
\caption{Phase-encode rows sampled by every static mask, including the relaxed row profile and the energy oracle ($f = 0$ is DC).}
\label{fig:mri-rows-all}
\end{figure}

\begin{figure}[!htbp]
\centering
\includegraphics[width=\textwidth]{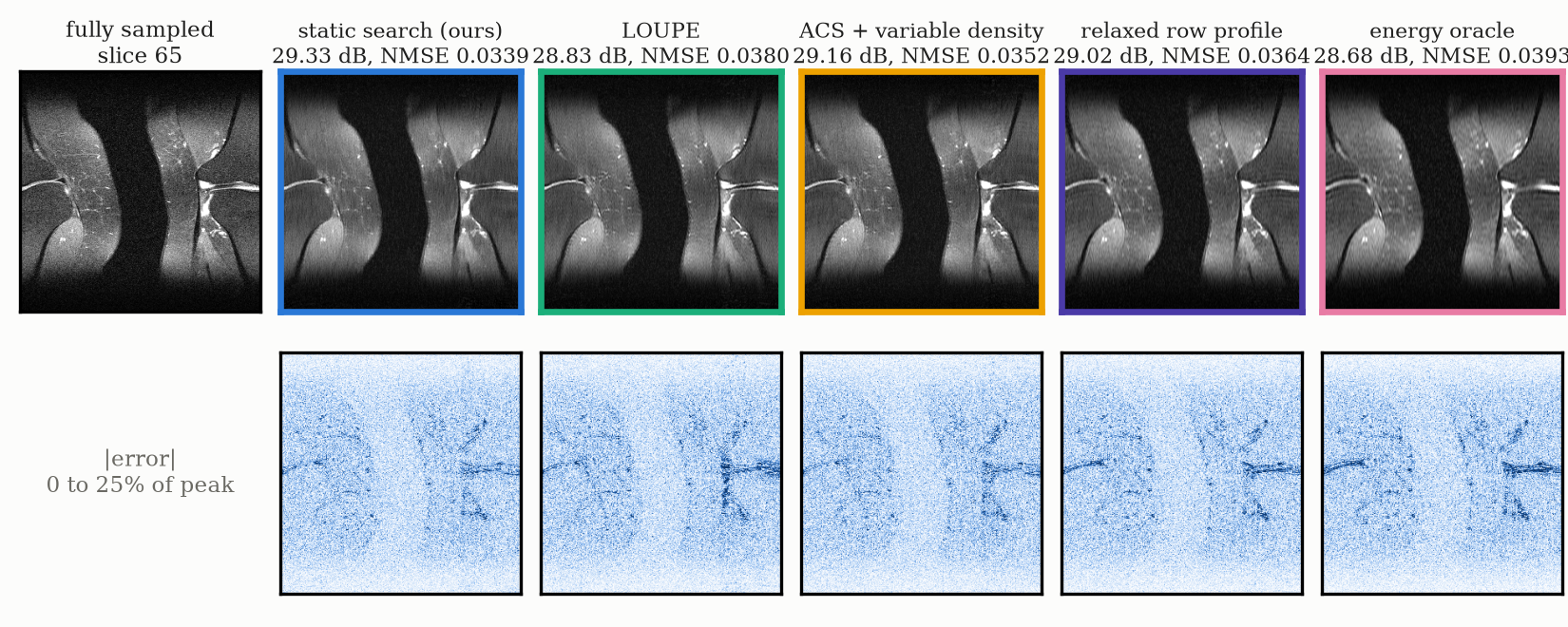}
\caption{A validation slice of 75th-percentile difficulty for ACS + variable density, at $s = 0.25$, reconstructed by the hold-out judge. Top: reconstructions. Bottom: absolute error $|$reconstruction $-$ fully sampled$|$, from 0 (white) to a quarter of the slice's 99.5th-percentile intensity (dark blue). PSNR and NMSE are per slice.}
\label{fig:mri-examples-p75}
\end{figure}

\paragraph{Static search.} Every run of the row-swap search stopped at a local optimum after $0$--$11$ accepted swaps; a little over half the gain ($57\%$ over all budgets) comes from starting at the best fixed variable-density draw rather than redrawing per slice, and the rest from the swaps. The static mask is the best of those tested on NMSE and PSNR at every budget (Figures~\ref{fig:mri-zf}, \ref{fig:mri-judge} and~\ref{fig:mri-examples}), and it also has lower NMSE than variable density under plain zero-filled reconstruction at every budget. SSIM agrees with NMSE against variable density at every budget, but center-concentrated masks score up to $1.0$ SSIM points ($\times 100$) higher at $s \ge 0.40$. Figure~\ref{fig:mri-rows-all} adds the relaxed row profile and the energy oracle to Figure~\ref{fig:mri-rows}; both concentrate their rows at low frequencies. Figure~\ref{fig:mri-examples-p75} shows a harder validation slice, of 75th-percentile difficulty: the search mask again has the lowest NMSE ($0.0339$, against $0.0352$ for variable density).

\paragraph{Relaxed row profile.} A relaxed row profile, 300 logits trained against U-Net A through a straight-through top-$k$ mask, leaves its energy initialization only with a small step ($2 \times 10^{-4}$; at $5 \times 10^{-3}$ no epoch improves on it at $s \ge 0.40$). Retuned, it matches variable density at $s = 0.40$ and $0.75$, improves on it only at $s = 0.50$ ($0.01527$ against $0.01546 \pm 0.00007$), loses at $s = 0.25$, and trails the search mask at every budget.

\paragraph{Sequential acquisition.} Starting from the ACS block, each of $R$ steps adds the unobserved rows where the coarse row-profile prior's predictive entropy is highest. Table~\ref{tab:mri-closed} gives every variant. Under the judge it loses to variable density at $s = 0.25$ and $0.40$ ($4.8\%$ and $5.5\%$ higher NMSE with $R = 16$), is level at $s = 0.50$, and never reaches the static search mask's NMSE. At $s = 0.75$ a single step is best ($1.8\%$ lower NMSE than variable density, level with the search mask), but that mask is the same for every slice: after the ACS block the coarse profile carries no slice-specific information, so the first step reads a static mask off the prior. Under zero-filled reconstruction it has lower NMSE than variable density at every budget, the pattern of Table~\ref{tab:recon-gain}. Ranking rows by the prior's expected energy instead of its entropy is worse at every budget and number of steps, so the entropy carries more than energy, but a prior over row energy does not tell the reconstructor what it is missing.

\paragraph{LOUPE.} We ported LOUPE \citep{bahadir2020loupe} to PyTorch (mask layers matching the original to $3 \times 10^{-7}$), restricted it to whole PE rows, and trained it until its row set stopped changing. Its mask is $15.5$--$31.0\%$ worse than ours under the judge, and worse than every heuristic at every budget, although it trains its reconstructor jointly. Its forward model assumes Hermitian k-space: at $s \le 0.50$ its masks hold only $7$--$8$ mirrored $\pm f$ row pairs (ours: $17$--$57$) and at $s = 0.75$ sample one half of k-space.

\begin{table}[!htbp]
\centering
\caption{Hold-out val NMSE under the judge U-Net B for sequential row acquisition with the coarse row-profile prior: starting from the ACS block, rows are added in 1, 4 or 16 steps, ranked by the prior's predictive entropy or, as a control, by its expected energy. ACS + variable density is the mean over 5 seeds. Bold: the lowest NMSE per budget, and any entry within two standard deviations of it, measured by the seed spread of ACS + variable density.}
\label{tab:mri-closed}
\small
\setlength{\tabcolsep}{4pt}
\begin{tabular}{lcccc}
\hline
\textbf{Mask} & $s = 0.25$ & $s = 0.40$ & $s = 0.50$ & $s = 0.75$ \\
\hline
ACS + variable density & 0.02471 & 0.01889 & 0.01546 & 0.00764 \\
static search (ours) & \textbf{0.02389} & \textbf{0.01860} & \textbf{0.01523} & \textbf{0.00751} \\
sequential acquisition, entropy, 1 step & 0.03031 & 0.02571 & 0.02061 & \textbf{0.00751} \\
sequential acquisition, entropy, 4 steps & 0.02670 & 0.02254 & 0.01916 & 0.00941 \\
sequential acquisition, entropy, 16 steps & 0.02589 & 0.01992 & 0.01559 & 0.00807 \\
control, expected energy, 1 step & 0.03359 & 0.02928 & 0.02381 & 0.01145 \\
control, expected energy, 4 steps & 0.02982 & 0.02438 & 0.01940 & 0.01209 \\
control, expected energy, 16 steps & 0.02699 & 0.02122 & 0.01790 & 0.00966 \\
\hline
\end{tabular}
\end{table}

\subsection{What the mask state adds}
\label{app:token}

To measure what the separate mask state of Section~\ref{sec:formulation} adds, we fine-tune two MNIST priors from the same D3PM with the same recipe ($60$ epochs, learning rate $2 \times 10^{-4}$, batch $512$): one with the mask state, and a control that keeps $0$ as both black and unobserved. The pair is repeated for $5$ fine-tuning seeds. Both priors drive the same acquisition, a random probe of $20\%$ of the budget followed by one greedy step on the prior's entropy, on the same $2560$ test images with the same probes ($5$ evaluation seeds). With the mask state the prior makes $10$--$15\%$ fewer errors on average (Table~\ref{tab:token}), in every fine-tuning seed at $s \le 0.3$ and in $4$ of $5$ at $s = 0.4$. The spread comes from the fine-tunes, not the evaluation, whose standard deviation over evaluation seeds is below $0.15$ errors; it grows with the budget as the error counts shrink.

\begin{table}[!htbp]
\centering
\caption{MNIST errors per test image with and without the separate mask state, probe + one greedy step, mean $\pm$ standard deviation over $5$ fine-tuning seeds. The last column counts the seeds in which the mask state has fewer errors.}
\label{tab:token}
\small
\setlength{\tabcolsep}{4pt}
\begin{tabular}{ccccc}
\hline
$s$ & \textbf{Without} & \textbf{With mask state} & \textbf{Fewer errors} & \textbf{Seeds} \\
\hline
0.10 & $28.72 \pm 0.36$ & $\mathbf{25.73} \pm 0.44$ & $10.4 \pm 0.9\%$ & 5/5 \\
0.20 & $10.08 \pm 0.22$ & $\mathbf{8.83} \pm 0.33$ & $12.4 \pm 2.8\%$ & 5/5 \\
0.25 & $4.98 \pm 0.14$ & $\mathbf{4.28} \pm 0.25$ & $14.2 \pm 4.5\%$ & 5/5 \\
0.30 & $2.07 \pm 0.08$ & $\mathbf{1.76} \pm 0.15$ & $15.0 \pm 7.4\%$ & 5/5 \\
0.40 & $0.22 \pm 0.01$ & $\mathbf{0.19} \pm 0.02$ & $14.4 \pm 10.8\%$ & 4/5 \\
\hline
\end{tabular}
\end{table}

\end{document}